\documentclass[a4paper,aps,pre,floatfix,nofootinbib,twocolumn,showpacs,bibtex]{revtex4}

\usepackage[dvips]{epsfig}
\usepackage[dvips]{graphicx}
\usepackage{latexsym,amsmath,verbatim}
\usepackage{color}

\newcommand{\av}[1]{\langle {#1} \rangle}

\newcommand{\ep}{\varepsilon}
\newcommand{\vp}{\phi}
\newcommand{\be}{\begin{equation}}
\newcommand{\ee}{\end{equation}}

\begin{document}

\title{The non-linear $q$-voter model}

\author{Claudio Castellano}

\affiliation{SMC, INFM-CNR and Dipartimento di Fisica, ``Sapienza''
  Universit\`a di Roma, P.le Aldo Moro 2, I-00185 Roma, Italy}

\author{Miguel A. Mu\~noz}

\affiliation{Departamento de Electromagnetismo y F\'\i sica de la
  Materia and Instituto de F\'\i sica Te\'orica y Computacional
  Carlos I, Facultad de Ciencias, Universidad de Granada, 18071
  Granada, Spain }

\author{Romualdo Pastor-Satorras}

\affiliation{Departament de F\'\i sica i Enginyeria Nuclear,
  Universitat Polit\`ecnica de Catalunya, Campus Nord B4, 08034
  Barcelona, Spain}

\date{\today}

\begin{abstract}
  We introduce a non-linear variant of the voter model, {\it the
    $q$-voter model}, in which $q$ neighbors (with possible
  repetition) are consulted for a voter to change opinion. If the $q$
  neighbors agree, the voter takes their opinion; if they do not have
  an unanimous opinion, still a voter can flip its state with
  probability $\ep$.  We solve the model on a fully connected network
  (i.e. in mean-field) and compute the exit probability as well as the
  average time to reach consensus by employing the backwards
  Fokker-Planck formalism and scaling arguments.  We analyze the
  results in the perspective of a recently proposed Langevin equation
  aimed at describing generic phase transitions in systems with two
  ($Z_2$ symmetric) absorbing states.  In particular, by deriving
  explicitly the coefficients of such a Langevin equation as a
  function of the microscopic flipping probabilities, we find that in
  mean-field the $q$-voter model exhibits a disordered phase for high
  $\ep$ and an ordered one for low $\ep$ with three possible ways to
  go from one to the other: (i) a unique (generalized voter-like)
  transition, (ii) a series of two consecutive transitions, one
  (Ising-like) in which the $Z_2$ symmetry is broken and a separate
  one (in the directed percolation class) in which the system falls
  into an absorbing state, and (iii) a series of two transitions,
  including an intermediate regime in which the final state depends on
  initial conditions.  This third (so far unexplored) scenario, in
  which a new type of ordering dynamics emerges, is rationalized and
  found to be specific of mean-field, i.e. fluctuations are explicitly
  shown to wash it out in spatially extended systems.
\end{abstract}

\pacs{05.70.Ln, 02.50.2r, 05.50.+q, 64.60.Ht}

\maketitle

\section{Introduction}

In a situation where one has to choose between two alternatives that
appear equally agreeable, a rather common way to remove the
uncertainty is to copy what somebody else (randomly selected among
acquaintances) does.  The voter model dynamics~\cite{Clifford73}
describes precisely this situation: Agents placed on the vertices of a
graph are characterized by a binary (spin) variable $\pm 1$; at each
time step two nearest neighbor vertices are selected and the first
copies the state of the second.  This can equivalently be expressed by
the flipping probability $f(x)$: A vertex with a fraction $x$ of
disagreeing neighbors has a linear probability $f(x)=x$ to flip.  The
iteration of this simple (parameter free) rule gives rise to
nontrivial ordering phenomena, that have drawn the attention of many
scholars, both in physics~\cite{Castellano09} and
beyond~\cite{nowak2006ed}.  From the point of view of statistical
physics, voter dynamics stands out as one of the very few
nonequilibrium processes amenable of exact analytical treatment in any
dimension~\cite{Frachebourg96}.  In more physical terms, it owes its
special character to the absence of surface tension~\cite{Dornic01}.
Contrary to the more common curvature-driven dynamics~\cite{Bray94},
in voter dynamics curved interfaces do not tend to reduce their
curvature and assume a straight shape.  This induces a slow domain
growth characterized, in two dimensions, by a logarithmic decay of the
density of active links (i.e. links connecting sites with opposite
opinion states).

A natural and relevant question, that has attracted interest in the
past years, has to do with how generic the voter behavior is. Early
work~\cite{DeOliveira93,Drouffe99,Molofsky99} already pointed out that
small changes in the dynamics destroy the voter behavior in two
dimensions. The voter model turns out to sit at the transition between
a ferromagnetic and a paramagnetic phase but it is only a point in a
generalized parameter space and any perturbation leads to a
drastically different behavior.  These and other results pointing
towards the fragility of the voter
behavior~\cite{Baronchelli06,Castello06,Dallasta07,Schweitzer09} raise
the question of whether the voter model is a peculiar exception or the
representative of a more generic class of models.

An answer to this question was provided by Dornic \textit{et
  al.}~\cite{Dornic01}.  The authors of this work showed explicitly
the existence of models, different from the pure linear voter, that
nevertheless exhibit its typical dynamical features. This led them to
conjecture that there is a proper generalized voter (GV) universality
class encompassing systems at ``an order-disorder transition driven by
the interfacial noise between two absorbing states possessing
equivalent dynamical roles, this symmetry being enforced either by
$Z_2$ symmetry of the local rules, or by the global conservation of
the magnetization''~\cite{Dornic01}.

Further progress in the understanding of this issue has been made by
Al Hammal \textit{et al.}~\cite{AlHammal05}, who worked out a generic
Langevin equation \cite{Gardinerbook} for critical phenomena with two
symmetric absorbing states and identified conditions for having a
transition from order to disorder belonging to the GV class.  Note
that in voter-like models there are two different competing phenomena:
One is the breaking of the $Z_2$ symmetry and the other one is the
possibility for the system to get trapped into an absorbing state. If
both occur in unison then the transition point is in the GV class.
Instead, if they occur separately, the $Z_2$ symmetry is broken first
(i.e. an Ising like transition occurs, and the system changes from
paramagnetic to ferromagnetic) and afterwards the system falls into
the corresponding absorbing state (i.e a directed percolation like
transition)~\cite{AlHammal05}. In this sense, the GV class can be
rationalized as the superposition of Ising and directed percolation
phase transitions.  The fact that the voter transition can be split
into two different ones was first reported in \cite{Drofeli}.

The picture devised in Ref.~\cite{AlHammal05} on the basis of generic
symmetry arguments has been recently substantiated by V\'azquez and
L\'opez~\cite{Vazquez08}. Starting from the microscopic spin dynamics
of a {\it non-linear voter model}, they have derived an explicit
Langevin equation for the magnetization, that coincides with the one
conjectured in Ref.~\cite{AlHammal05}.  In this way, it is possible to
precisely determine, depending on the analytical form of the
microscopic flipping probability $f(x)$, which of the two scenarios
above occurs.

In this paper we provide an assessment of the picture presented in
Refs.~\cite{AlHammal05,Vazquez08} by proposing a microscopically
motivated non-linear voter model and analyzing it, both at the
mean-field level and numerically.  The model we consider, the
$q$-voter model, represents a simple generalization of the original
voter model: each individual interacts with a set of $q$ of his
nearest neighbors; if all $q$ neighbors share the same state, the
individual conforms to this state.  Otherwise, if the $q$ neighbors do
not agree, the individual flips with a probability $\ep$.  The
$q$-voter model is directly inspired in models of ordering dynamics in
which each update step involves more than two individuals
\cite{Lambiotte07,Sznajd00,hegselmann02,Drofeli,galam05:_local,krapivsky2003dmr},
and represents a practical and simple way to introduce
nonlinearity in the voter dynamics at a microscopic level.

We study the model phenomenology analytically via a mean-field approach, and
numerically in two dimensions, uncovering a rich phenomenology.
Depending on the value of $q$, the model exhibits all the possible
transitions of a system with two symmetric absorbing states, as
described above.  Interestingly, at mean-field level the voter
behavior is restricted only to very specific values of $q$ ($q=2$ and
$q=3$), two separated phase transitions occur in $2 < q <3$, and,
otherwise ($q<2$ and $q>3$) we find apparently novel phenomenology
(i.e. dependence on the initial conditions and a double transition of
a different type). Direct numerical simulations of the model on a
fully connected network are presented to back up the mean-field
results. On the other hand, in a $d=2$ lattice, we recover the picture
presented in Ref.~\cite{AlHammal05}, with a single voter-like critical
point.

The paper is structured as follows: After Section~\ref{definition},
where the model is defined at a microscopic level, in
Sec.~\ref{mean-field} we perform a mean-field analysis, by means of
both a Fokker-Planck and a Langevin approach; a more detailed study of
the case $q=4$ is also presented.  Results for finite dimensional
systems are presented in Sec.~\ref{finitedimensions}. Finally, in
Sec.~\ref{conclusions} we  discuss  our findings.

\section{Definition of the $q$-voter model}
\label{definition}

We consider a non-linear voter model defined on a lattice (or network)
of $N$ sites. Each site hosts a spin, with value $\pm 1$. The dynamics
is given by the following update rule:
\begin{itemize}
\item At a given time $t$, choose one spin at random, located at site
  $i$.
\item Choose at random $q$ neighbors of site $i$. In order to
  simplify the numerical analysis, and allow for an arbitrary value of
  $q$ in regular lattices, we consider here the possibility of
  repetition, i.e. a given neighbor can be selected more than once.
\item If all the $q$ neighbors are in the same state, the original
  spin takes the value of the $q$ neighbors.
\item Otherwise, if the $q$ neighbors are different, the original
  spin flips with probability  $\ep$.
\item Time is updated $t \to t + 1/N$.
\end{itemize}
It is easy to see that this model is non-linear. Consider the
probability that a site flips as a function of the fraction $x$ of
disagreeing neighbors, that is,
\begin{equation}
  f(x,q) = x^q + \ep \left[1-x^q - \left(1-x\right)^q\right].
  \label{f}
\end{equation}
Notice that, although in the original definition of the model $q$ is
an integer, Eq.~(\ref{f}) makes sense for any $q>0$ and can therefore
be considered as the definition of the $q$-voter model for real values
of $q$.  For $q=1$, we recover, for any value of $\ep$, the standard
voter model, namely $f(x,1)= x$.  Non-linear behavior arises for $q \ne
1$.  Observe also that if the fraction of disagreeing neighbors
vanishes, i.e.  $x=0$, the configuration is absorbing, $f(0,q)=0$.

The $q$-voter model bears some resemblance to other opinion dynamics
models introduced recently in the literature. For example, the
``vacillating voter'' model~\cite{Lambiotte07} is very similar to the
$q$-voter with $q=2$ and $\ep=1$, apart from the possibility to select
twice the same neighbor (repetition).  The case with $q=2$ and $\ep=0$
is instead similar to the Sznajd model~\cite{Sznajd00}, in its
formulation where a pair of agreeing agents convince only one of their
neighbors~\cite{Slanina03}.

\section{Mean-field theory}
\label{mean-field}

In order to gain an understanding of the $q$-voter model behavior, it
is useful to consider it first at the mean-field level (that is, on a
fully connected network), for which several analytical tools have been
recently developed.

\subsection{Backward Fokker-Planck approach}
\label{sec:backw-fokk-planck}

Following Refs.~\cite{Sood05,Antal06,Sood08}, we can study the
mean-field theory of the $q$-voter model by applying the backward
Fokker-Planck (BFP) technique~\cite{Gardinerbook}. Consider a time
$t$, in which there are $n$ spins in state $+1$. The state of the
system is fully defined by this quantity, plus the transition rates to
go to a state with $n \pm1$ spins in state $+1$. Denoting this
transition probabilities by $p_{n\pm1,n}$, then
\begin{eqnarray}
  p_{n+1,n} &=& (1-x)  f(x,q)\\
  p_{n-1,n} &=& x  f(1-x,q)\\
  p_{n,n} &=& 1 - p_{n+1,n} -p_{n-1,n},
\end{eqnarray}
all the rest of values of $p_{n',n}$ being equal to zero.  Here we
consider $x=n/N$ as the probability of selecting a $+1$ spin when a
vertex in randomly chosen, a simplification which provides valid
results in the limit of large $N$.
With this definition there are two absorbing states, $n=0$ and $n=N$
(i.e. $x=0$ and $x=1$).

The quantity $n$ performs in time a biased one dimensional random
walk, between two absorbing states. The random walk is fully defined
in terms of a backward master equation, taking the form
\cite{Gardinerbook}
\begin{eqnarray}
& &  \frac{\partial P(n, t| n', t')}{\partial t'} \nonumber \\
&=& T(n'+1|n') [P(n, t| n'+1, t') - P(n,t| n', t')] \nonumber\\
&+& T(n'-1|n') [P(n, t| n'-1, t') - P(n,t| n', t')],
  \label{eq:16}
\end{eqnarray}
where $P(n,t| n', t')$ is the probability of having $n$ spins $+1$ at
time $t$, provided there were $n'$ at time $t' \leq t$.
Eq.~(\ref{eq:16}) is given in terms of the transition rates (per unit
time)
\begin{equation}
  T(n|n') = \frac{p_{n,n'}}{\Delta},
\end{equation}
with $\Delta=1/N$. The master equation can be transformed, via a
diffusion approximation, into a BFP equation for the reduced variable
$x=n/N$, by expanding Eq.~(\ref{eq:16}) up to second order in
$\Delta$. In this expansion, the BFP equation takes the form
\begin{eqnarray}
\frac{\partial P(x, t| x', t')}{\partial t'} &=&
 v(x')  \frac{\partial P(x, t| x', t')}{\partial x'} \nonumber
\\
&+&
 \frac{1}{2}
D(x')  \frac{\partial^2 P(x, t| x', t')}{\partial x'^2},
\label{eq:3}
\end{eqnarray}
with a drift 
\begin{eqnarray}
  v(x) &=& \Delta [T(n+1|n)-T(n-1|n)] \nonumber\\ 
       &=& (1-x) f(x,q) - x f(1-x,q),
\label{drift}
\end{eqnarray}
and a diffusion coefficient
\begin{eqnarray}
  D(x) &=& \Delta^2 [T(n+1|n)+T(n-1|n)] \nonumber\\
       &=& \frac{1}{N} \left[(1-x) f(x,q) + x f(1-x,q)\right].
\end{eqnarray}
For the generic BFP equation Eq.~(\ref{eq:3}), the exit probability
$E(x)$, i.e. the probability that, starting from an initial density
$x$ of $+1$ vertices, the absorbing state $+1$ is reached, satisfies
the differential equation \cite{Gardinerbook}
\begin{equation}
  v(x) \partial_x E(x) + \frac{1}{2}D(x) \partial^2_x E(x) =
  0, \label{eq:1} 
\end{equation}
subject to the boundary conditions $E(0)=0$ and $E(1)=1$, while the
average time until consensus, $T(N, x)$ is given by
\cite{Gardinerbook}
\begin{equation}
   v(x) \partial_x T(N,x)  + \frac{1}{2} D(x)  \partial^2_x T(N,x) = 
   -1,\label{eq:6} 
\end{equation}
with boundary conditions $T(N,0)= T(N,1)=0$.

The standard voter model in Eq.~(\ref{eq:16}) corresponds to $q=1$,
for which we find $v(x)=0$ and $D(x)=2x(1-x)/N$. This leads to
$E(x)=x$ and $T(N,x) = - N \left[ x \ln x + (1-x)\ln (1-x)\right]$;
thus $T(N,1/2) \sim N$ and $T(N,1/N) \sim \ln N$
\cite{Sood05}. Moreover, it is easy to see from Eqs.~(\ref{eq:1}) and
(\ref{eq:6}), that the condition $v(x)=0$ is necessary and sufficient
to yield, for any diffusion $D(x)$ (as long as $D(x) \propto
\frac{1}{N}$), $E(x)=x$ and $T(N,1/2) \sim N$, that are the two main
signatures of voter behavior in mean-field.

In order to have mean-field voter behavior, we must then consider the
cases in which the drift $v(x)$ vanishes. Let's look at the different
possibilities:
\begin{itemize}
\item
 For $q=2$,
\begin{equation}
  v(x) = (-1 + 2 \ep) (1 - x) x (1 - 2 x).
\end{equation}
Therefore, $\ep=1/2$ leads to voter behavior. But for $\ep=1/2$,
$f(x,2)=x$, so that in this case the $q$-model {\em coincides} with
the usual voter model.
\item For $q=3$,
\begin{equation}
  v(x) = (-1 + 3 \ep) (1 - x) x (1 - 2 x).
\end{equation}
Again, $\ep=1/3$ leads to zero drift and hence to voter behavior.
However, in this case $f(x,3)=x^3-x^2+x \ne x$, so that the $3$-voter
model is a case belonging nontrivially to the GV class.
\item For $q=4$ instead, 
\begin{eqnarray}
  v(x) &=& (1 - x) x (1 - 2 x) \nonumber \\
  &\times& [-1 + 4 \ep +x(1-x)(1-2\ep)].
\end{eqnarray}
No value of $\ep$ can cancel the drift; therefore, voter behavior is,
in principle, not possible.
\end{itemize}

\subsection{Langevin equation approach}
\label{sec:lang-equat-appr}

Further understanding is provided by applying the formalism developed
in Refs.~\cite{AlHammal05,Vazquez08}. In this approach one focuses on
the magnetization  $\vp = 2 x -1$. In this variable,
the drift takes the form, at lowest level in powers of $\vp$,
\begin{equation}
  v(\vp) = (1-\vp^2) \left( a \vp - b    \vp^3 \right).
  \label{eq:2}
\end{equation}
This corresponds to the usual terms in a continuous description for
systems with a $Z_2$ symmetry (i.e in the Ising class \cite{HH}),
multiplied by a factor $(1-\vp^2)$ imposing the existence of two
absorbing states.  The drift can be written as derived from a
potential: $v(\vp) = - d V(\vp)/ d \vp$, i.e.
\begin{equation}
  V(\vp) = -\frac{a}{2} \vp^2 + \frac{a+b}{4} \vp^4 - \frac{b}{6}\vp^6.
\label{V}
\end{equation}

\begin{figure}
  \begin{center}
    \includegraphics*[width=6cm]{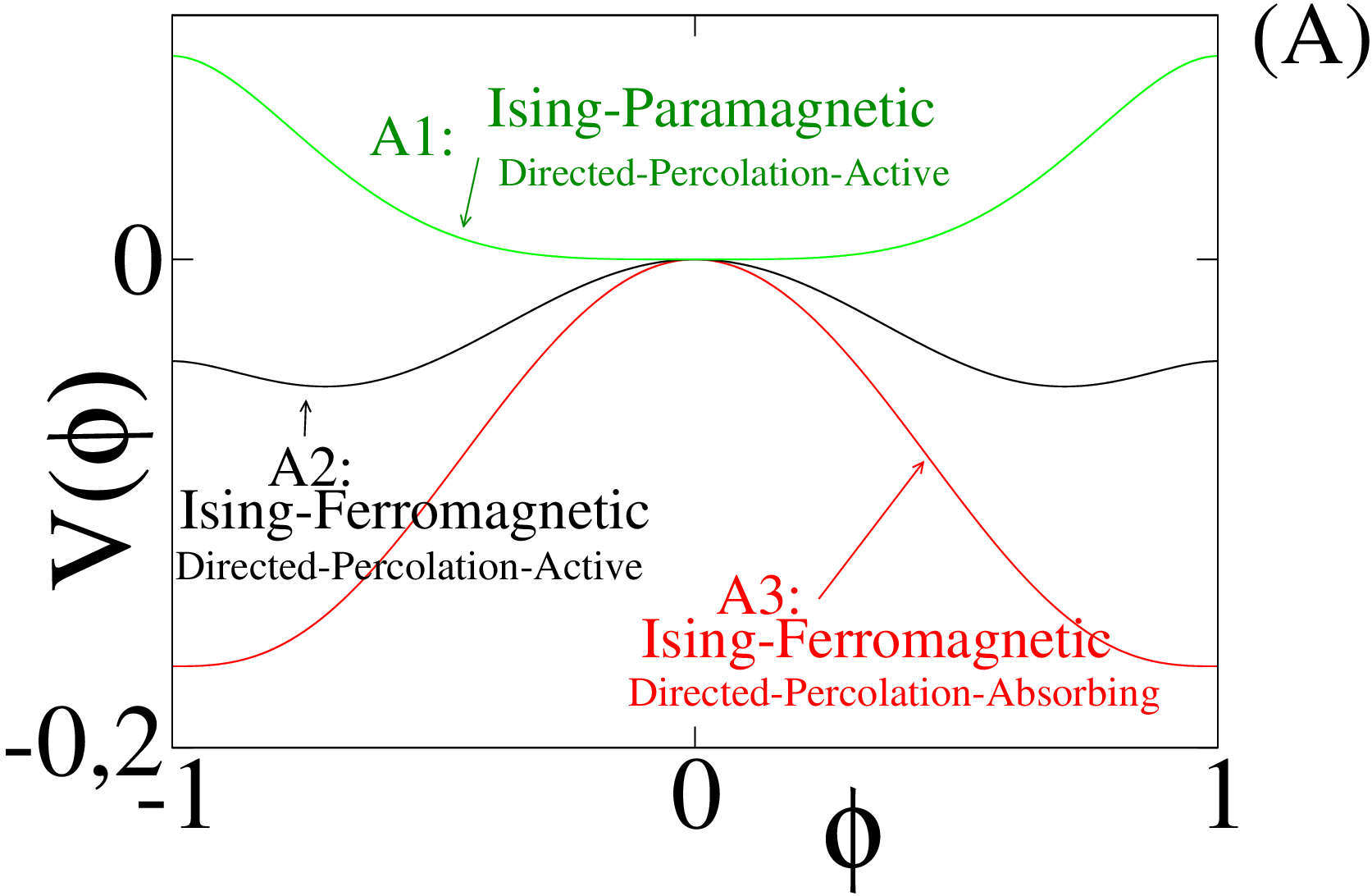}
    \includegraphics*[width=6cm]{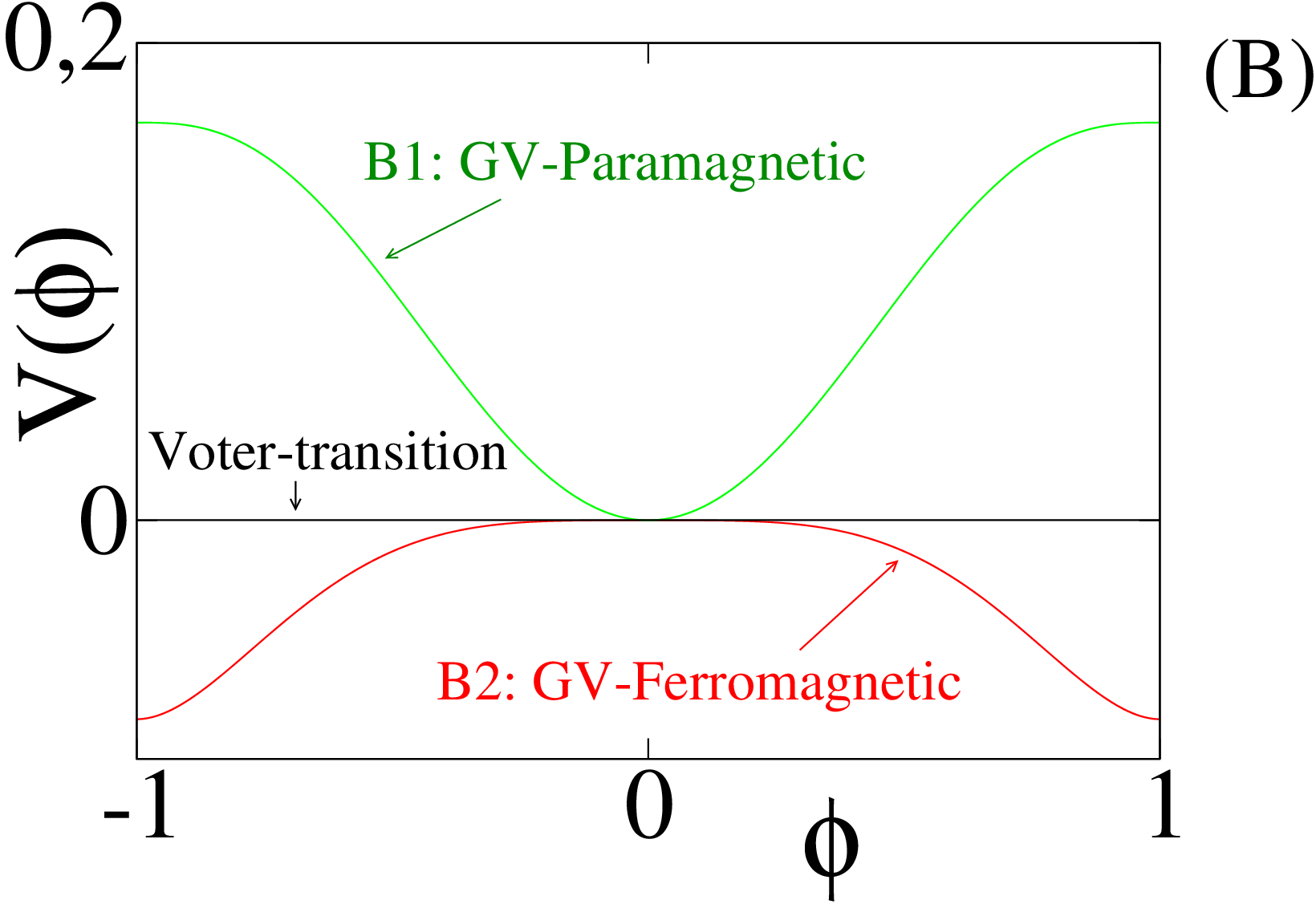}
    \includegraphics*[width=6cm]{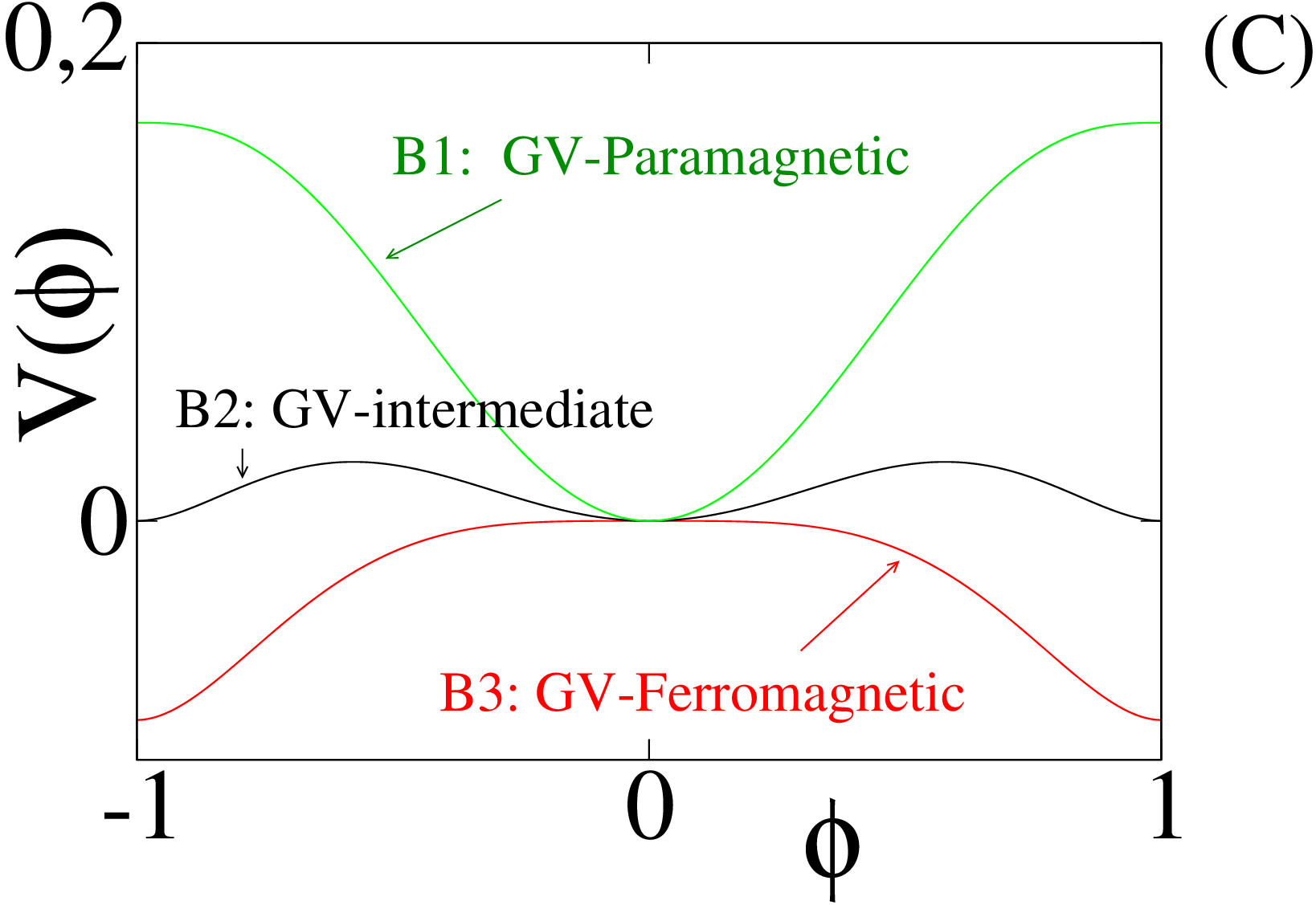}
  \end{center}
  \caption{(Color online) Potential $V(\phi)$, as defined by
    Eq.~(\ref{V}) for $b>0$ (a), $b=0$ (b), and $b<0$ (b).}
  \label{potentials}
\end{figure}

This function has $5$ extrema, obtained from the condition $v(\vp)
=0$, which are $\vp = 0$, $\vp = \pm1$ and $\vp = \pm
\sqrt{\frac{a}{b}}$.  Their role is clarified by the concavity of
$V(\vp)$, that turns out to be
\begin{eqnarray}
  V''(0) &=& -a,\\
  V''(\pm1) &=& 2(a-b),\\
  V''\left(\pm\sqrt{\frac{a}{b}}\right) &=& 2a\left(1-\frac{a}{b}\right).
\end{eqnarray}
The extrema at $0$ (origin) and $\pm1$ (absorbing barriers) are always
relevant.  The extrema at $\pm \sqrt{\frac{a}{b}}$ make physical sense only
when $0 < \frac{a}{b} < 1$, otherwise they are imaginary or
non-accessible. According to the interpretation in Ref.~\cite{AlHammal05},
there are the following possible scenarios, depending on $b$ (see
Fig.~\ref{potentials}):
\begin{itemize}
\item For $b>0$, if $a<0$ the system is paramagnetic; an Ising transition
  occurs for $a=0$ and (afterwards) an absorbing-state (directed percolation)
  transition takes place at $a=b$ (see Fig.~\ref{potentials}a).

\item The case $b=0$ corresponds to the voter case in which the
  potential identically vanishes at the transition point $a=0$ (see
  Fig.~\ref{potentials}b).

\item For $b < 0$, if $a$ is very negative the system is paramagnetic; then,
  at a negative value, $a=b$, a pair of symmetric new minima appear at $\pm 1$
  (this generates an ``intermediate phase'' with three competing minima). At
  $a=0$ the stability of the origin changes (see Fig.~\ref{potentials}c),
  and only the minima at $\pm 1$ remain. 
\end{itemize}

In Ref.~\cite{AlHammal05} it was argued that the intermediate phase
appearing for $b<0$ is absent in spatially extended systems:
fluctuations wash it away, and the central curve in
Fig.~\ref{potentials}c becomes as the lowest one.  The reason for this
is simple: As soon as minima at the absorbing barriers appear,
fluctuations become asymmetric, i.e. they can take the system from the
minimum at the origin to the barriers, but not the other way around.
Therefore, in a ``renormalized'' picture the third case is argued to
coincide with the $b=0$ case, and thus lead to a unique GV transition.
Hence, only two scenarios are expected to exist in the presence of
fluctuations.

The coefficients $a$ and $b$ can be explicitly computed for the
$q$-voter by combining Eqs.~(\ref{eq:2}), (\ref{drift}), and (\ref{f})
for generic values of $q$ and $\ep$. This leads to
\begin{eqnarray}
  a &=& 2^{-q+1} (q-1)-2\ep( 1-2^{-q+1}), \label{eqa} \\
  b &=& 2^{-q} (q-1)(q-2) \left(1-\frac{q}{3} \right)\nonumber \\
  &+&2 \ep [1-2^{-q}(2-q+q^2)].\label{eqb}
\end{eqnarray}
Before discussing what occurs for generic values of $q$, we remark that the
coefficient $b$ can be made to vanish identically only for $q=1$, $q=2$, and
$q=3$, while the coefficient $a$ vanishes for any $\ep$ only if $q=1$. Hence,
we recover the previous results in Sec.~\ref{sec:backw-fokk-planck}.  For
$q=1$ the potential vanishes; the $q$-voter model coincides with the usual
voter ($\ep$ plays no role and the system sits at a critical point).  For
$q=2$ one has $b=0$ and $a=1/2-\ep$: the potential is exactly zero for
$\ep=1/2$ so that one recovers voter behavior at the transition between a
paramagnetic and a ferromagnetic phase.  For $q=3$, we have $b=0$ and
$a=1/2(1-3\ep)$: again the potential vanishes for $\ep=1/3$ (transition point)
and voter behavior is found, separating an ordered phase (for small $\ep$)
from a disordered one.

\begin{figure}
  \begin{center}
 \includegraphics*[width=7cm]{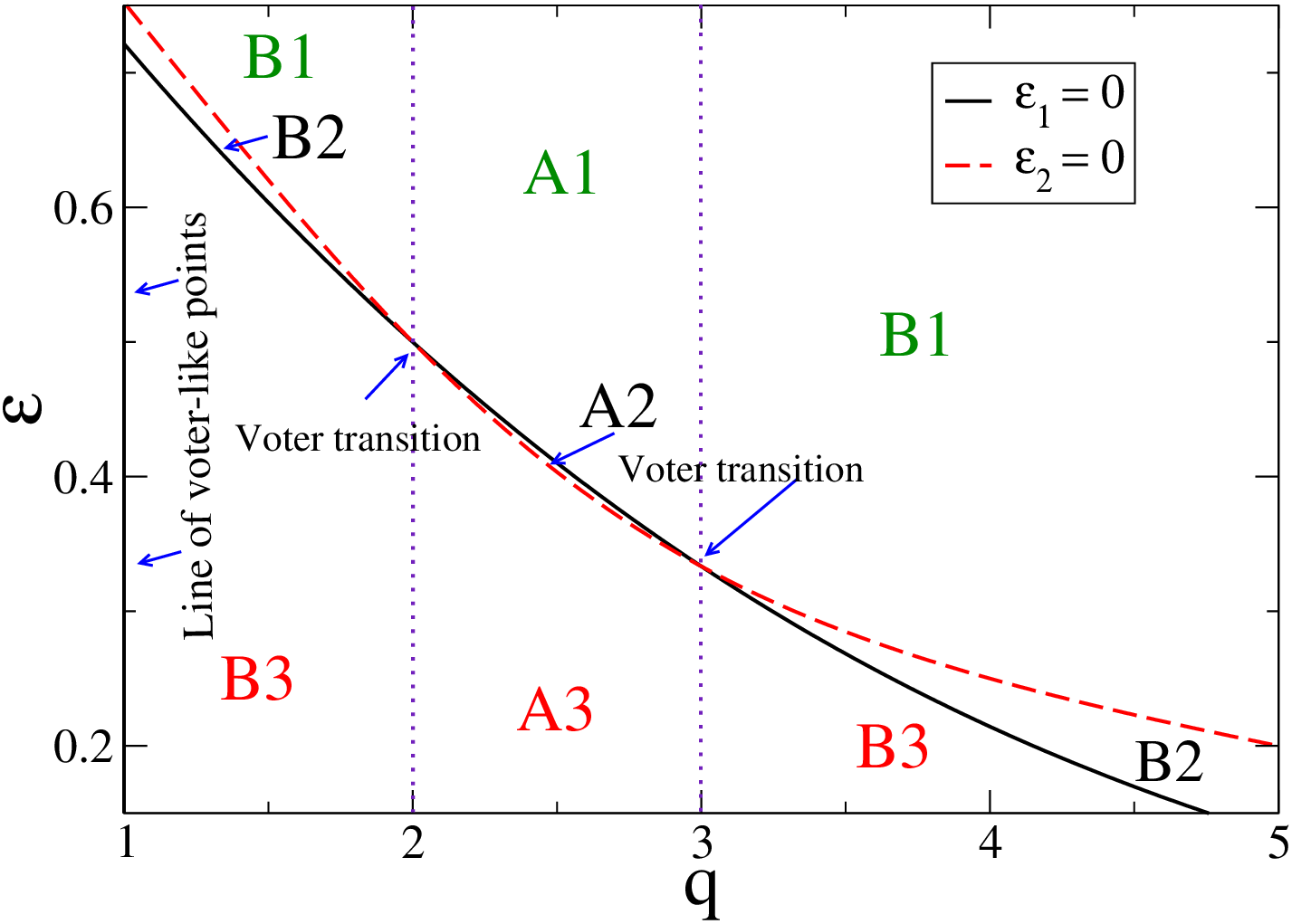}
\end{center}
\caption{(Color online) Mean field phase diagram of the $q$-voter
  model.  At $q=1$, $q=2$ and $q=3$ (marked with vertical lines) GV
  transitions occur.  For any other value of $q$ there are two
  different transitions: at $\ep_1=0$ (solid line) the up-down ($Z_2$)
  symmetry is broken, while at $\ep_2=0$ (dashed line) the barriers at
  $\pm 1$ become absorbing states. For $2<q<3$ there are an Ising
  transition followed by a directed percolation one. For $q<2$ and
  $q>3$ an intermediate phase exists.}
\label{Fases2}
\end{figure}

For analyzing what happens for generic values of $q$, it is useful to
calculate the boundaries of the interval of values of $\ep$ for which the
extrema in $\pm\sqrt{\frac{a}{b}}$ have physical values. We define as $\ep_1$
the value at which $a=0$ so that the extrema are in $0$. From
Eq.~(\ref{eqa}) we obtain
\be \ep_1=\frac{q-1}{2^q-2}.  \ee Instead, $\ep_2$ is the value for which the
stability at the origin changes, i.e. for which $a=b$. From
Eq.~(\ref{eqb}) we find
\be \ep_2=\frac{\frac{q^3}{3}-2q^2+\frac{17}{3}q-4}{2^{q+2}-2(4-q+q^2)} \ee
The behavior of $\ep_1$ and $\ep_2$ as a function of $q$ is plotted in
Fig.~\ref{Fases2}.
It is important also to notice that, for $q>1$, $a$ is smaller than zero for
large $\ep$ so that there is a paramagnetic phase above the black solid line
and a ferromagnetic one for small $\ep$.  This corresponds to intuition: $\ep$
plays the role of a sort of noise in the dynamics.  Instead, for $q<1$ (not
represented in Fig.~\ref{Fases2}) the situation is reversed: the paramagnetic
phase is now for small $\ep$ and the ferromagnetic one is for large values of
$\ep$.

The nature of the transition between the two phases varies depending on the
value of $q$. We have already established that  $q=1$ represents a
voter line, while for $q=2$, and
$ q=3$ (marked with vertical lines in Fig.~\ref{Fases2}) there is a
voter transition point for the appropriate values of $\ep$.

For $2<q<3$, $\ep_1$ is larger than $\ep_2$, and the
nature of the extrema is therefore as follows:
\begin{itemize}
\item For $\ep > \ep_1$, $0$ is a minimum and $\pm1$ are maxima.  The model is
  then in the paramagnetic phase (case A1 in Fig.~\ref{potentials}).

\item For $\ep_2 < \ep < \ep_1$, $0$ and $\pm1$ are maxima,
  and $\pm \sqrt{\frac{a}{b}}$ are minima.
The mode is in the ferromagnetic phase (case A2 in Fig.~\ref{potentials}).

\item For $\ep < \ep_2$, $0$ is a maximum and $\pm1$ are minima. The model is
  in the ferromagnetic absorbing phase (case A3 in Fig.~\ref{potentials}).

\end{itemize}
Obviously, in this case the double transition scenario described in
Ref.~\cite{AlHammal05} applies: Starting from large values of $\ep$,
first (at $\ep=\ep_1$) a transition in the Ising class, from a
paramagnet to a ferromagnet, occurs; then (at $\ep_1$) a transition of
directed percolation type appears and the system becomes fully
ordered. The same scenario occurs for $q<1$, where, as mentioned
above, the paramagnetic phase is for $\ep<\ep_1$ and the absorbing one
is for $\ep>\ep_2$.

On the other hand, for $ 1<q<2$ and $ q>3$, 
the relative positions of $\ep_1$
and $\ep_2$ are swapped, $\ep_1 < \ep_2$.  There is an intermediate interval
$\ep_1<\ep<\ep_2$ such that $0$ and $\pm 1$ are minima separated by maxima in
$\pm \sqrt{a/b}$.  The nature of the extrema is then as follows:
\begin{itemize}

\item For $\ep > \ep_2$, $0$ is a minimum and $\pm1$ are maxima.
  The model is in the paramagnetic phase 
(case B1 in Fig.~\ref{potentials}).
\item For $\ep_1 < \ep < \ep_2$, $0$ and $\pm1$ are minima,
  and $\pm \sqrt{\frac{a}{b}}$ are maxima
(case B2 in Fig.~\ref{potentials}). 
\item For $\ep < \ep_1$, $0$ is a maximum and $\pm1$ are
  minima. The model is then in the ferromagnetic absorbing phase
(case B3 in Fig.~\ref{potentials}).

\end{itemize}
In the intermediate interval the (mean-field) system exhibits
ferromagnetic or paramagnetic behavior depending on the initial
condition, with basins of attraction determined by the separatrices
$\pm \sqrt{a/b}$.  The transition is complicated and there is no voter
behavior at a mean-field level, as will be illustrated in the
forthcoming subsection.

\subsection{Analysis of the $q=4$ case}

\begin{figure}
  \begin{center}
    \includegraphics*[width=7cm]{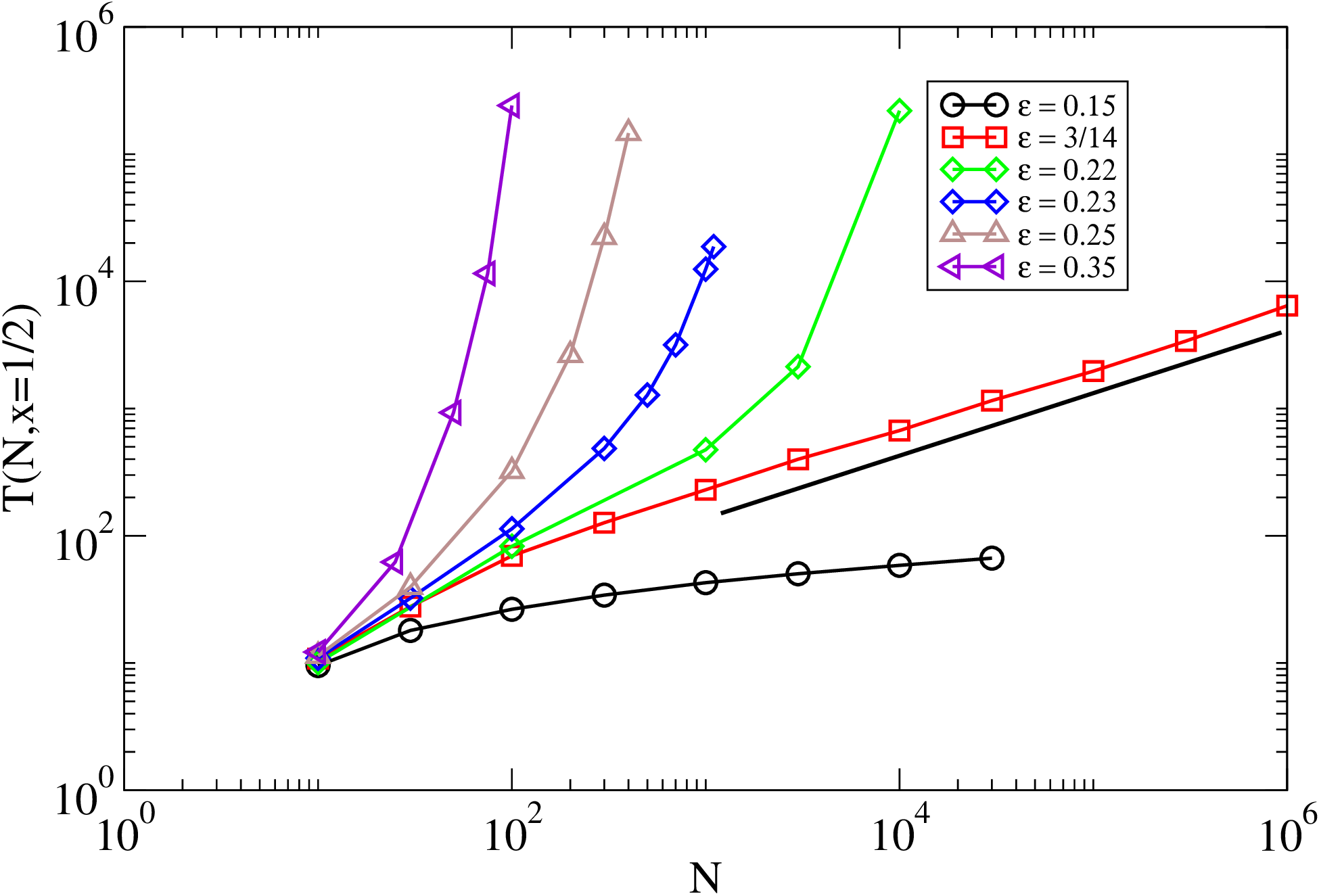}
  \end{center}
  \caption{(Color online) Average consensus time $T(N,x=1/2)$ as a
    function of $N$, for a fully connected system with $q=4$ and
    different values of $\ep$. The straight line has slope $1/2$. From
    bottom to top: $\epsilon=0.15, 3/14, 0.22, 0.23, 0.25$, and
    $0.35$.  }
  \label{TvsNq=4FC}
\end{figure}

The case $q=4$ is the smallest integer value of $q$ for which
the coefficients $a$ and $b$ cannot be made identically equal to zero
simultaneously:
\begin{equation}
  a= \frac{3-14 \ep}{8} \qquad \mathrm{and} \qquad b = \frac{2
    \ep-1}{8}. 
\end{equation}
Correspondingly we have $\ep_1=3/14$ and $\ep_2=1/4$.  In
Figure~\ref{TvsNq=4FC} we plot the average consensus time as a
function of $N$ for $q=4$ and several values of $\ep$, obtained by
numerical simulations of the $q$-voter model on fully connected
networks of different size.
 For $\ep>1/4$ the growth is exponential, as
expected in the paramagnetic phase.
For $\ep<3/14$ the growth is logarithmic, as expected in the
ferromagnetic phase.   For $3/14 < \ep <1/4$ there is a
crossover, but asymptotically the growth is exponential.  For
$\ep=3/14$ the growth is proportional to $N^{1/2}$, different from the
voter linear behavior.  

A simple scaling argument allows us to understand the growth law
$N^{1/2}$ for the average consensus time at the transition point
$\ep=3/14$.  For $q=4$ and $\ep=3/14$ the potential has parameters
$a=0$, $b=-1$, hence $V(\vp) = -\vp^4/4$ at leading order.  For
initial conditions $x=1/2$, $\vp=0$, so that at the beginning only
diffusion matters.  When $\vp$ becomes sufficiently large drift comes
in, the motion becomes ballistic and in an interval depending
logarithmically on $N$ consensus is reached.  How much time is spent
in the diffusion stage?  This interval lasts a time $t^*$ that is
estimated by equating the drift $v \sim \vp^3$ with the effective
velocity of the diffusion motion, $\sqrt{Dt^*}/t^* =
\sqrt{D/t^*}$. Hence $t^* = D/v^2 \sim D/\vp^6$.  During this time
interval the diffusive motion gives a displacement $\vp^* =
\sqrt{Dt^*}$ yielding $t^* = (\vp^*)^2/D$.  Equating the two
expressions gives $D \sim (\vp^*)^4$.  On the other hand the diffusion
coefficient $D$ is $D = (\Delta \vp)^2/(\Delta t) = (1/N)^2/(1/N) =
1/N$, from which $\vp^* \sim N^{-1/4}$.  Hence the total time to
consensus is \be T(N,1/2) \approx (\vp^*)^2/D + \log(N) \sim N^{1/2} +
\log(N) \sim N^{1/2}.  \ee

With respect to the exit probability $E(x)$, we can apply the
formalism in Sec.~\ref{sec:backw-fokk-planck}, solving
Eq.~(\ref{eq:1}) for the drift and diffusion functions 
\begin{eqnarray}
  v(x)&=& -\frac{1}{7} (1-x) x (1-2 x)^3,\\
  D(x)&=& \frac{1}{7N} (1-x) x \left(24 x^2-24 x+13\right),
\end{eqnarray}
to obtain, applying standard stochastic techniques \cite{Crowbook}, 
\begin{equation}
  E(x) = \frac{\int_0^x e^{-\frac{N}{12}  (2 y-1)^2}
    \left(24 y^2-24 y+13\right)^{7 N/72} \, dy}{\int_0^1
    e^{-\frac{N}{12}  (2 y-1)^2} \left(24 y^2-24
      y+13\right)^{7 N/72} \, dy}. 
  \label{eq:4}
\end{equation}
\begin{figure}
  \begin{center}
    \includegraphics*[width=7cm]{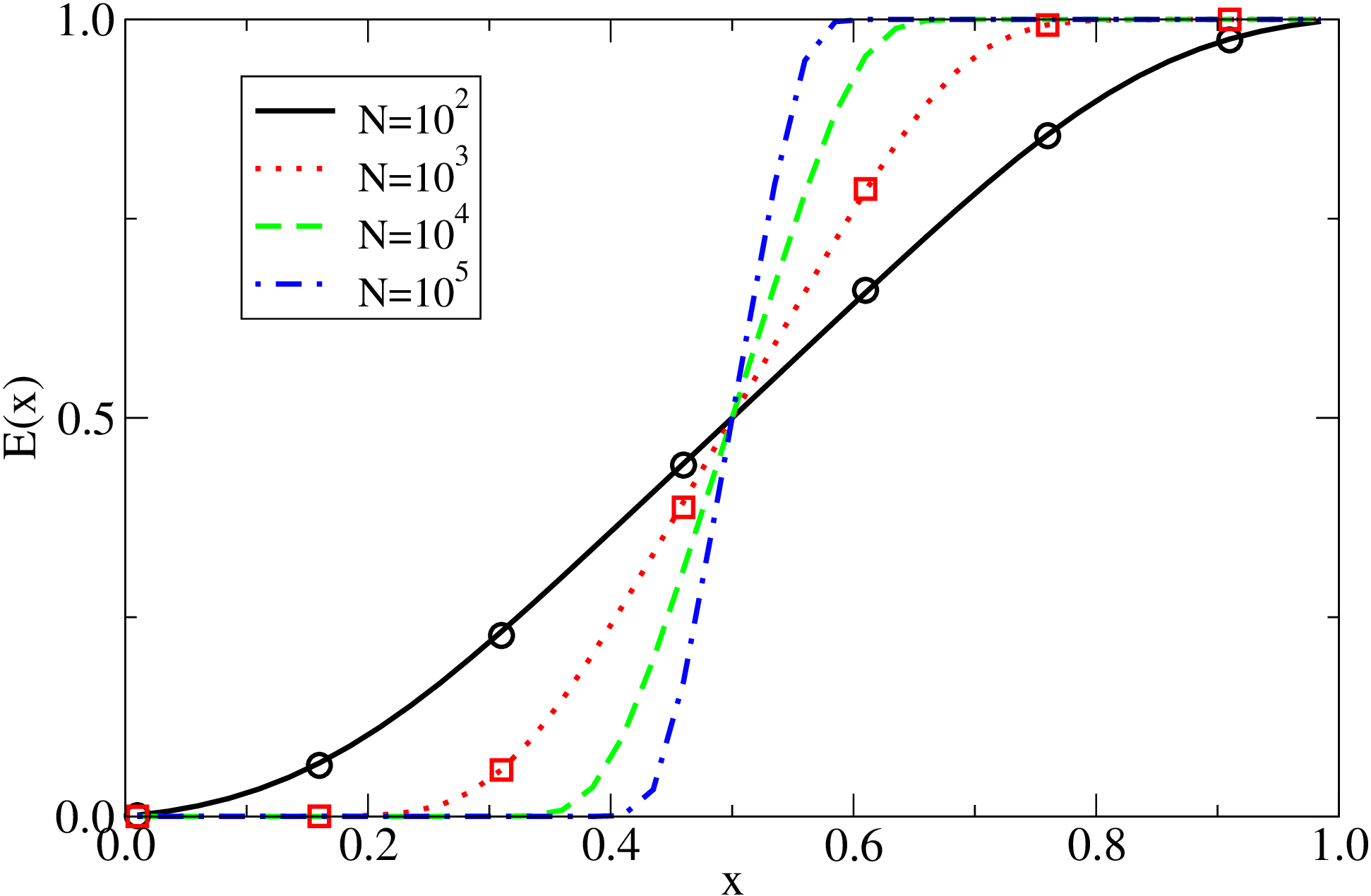}
  \end{center}
  \caption{(Color online) Exit probability $E(x)$ as a function of
    $x$, and different system sizes $N$, for a fully connected system
    with $q=4$ and $\ep=3/14$: the larger the size the steeper the
    slope at $x=1/2$.  Full lines represent the numerical integration
    of Eq.~(\ref{eq:4}); symbols stand for direct numerical
    simulations.}
  \label{Eq=4FC}
\end{figure}

In Fig.~\ref{Eq=4FC} we show the results of the numerical integration
of Eq.~(\ref{eq:4}) for different values of $N$ (solid lines),
together with results form direct numerical simulations of the
$q$-voter model in fully connected networks. It is apparent that
$E(x)$ tends to a step function at $x=1/2$ in the limit $N \to
\infty$.  No voter behavior is therefore observed in this mean-field case.

\section{Behavior in finite dimensions.}
\label{finitedimensions}

\subsection{$d=1$}
Even if in $d=1$ the number of different nearest neighbors (in a
square-lattice) is $2$, the parameter $q$ can be kept arbitrary.  It is
easy to see that the one-dimensional $q$-voter model can be mapped
onto the model of non-conservative voters recently introduced in
Ref.~\cite{Lambiotte08}.  In such a model, the relevant parameter
$\gamma$ is given by the ratio $p_2/p_1$, where $p_i$ is the flipping
probability for a site surrounded by $i$ disagreeing neighbors. In the
$q$-voter model $p_2$ trivially equals 1 for any $q$, while
$p_1=2^{-q}+\ep(1-2^{1-q})$. Hence 
\be 
\gamma =
\frac{2^q}{1+\ep(2^q-2)}. 
\ee 
This equation leads to the conclusion
that the value $\ep=1/2$ yields, for any $q$, a voter behavior
($\gamma=2$).  Analogously one can see that no $\ep$ can give the
value $\gamma=1$, implying the 'vacillating voter'
behavior~\cite{Lambiotte07}.  The mapping from $q$ and $\ep$ to
$\gamma$ allows us to deduce, from the results of
Ref.~\cite{Lambiotte08}, the nontrivial shape of the exit probability
$E(x)$ in $d=1$.

\subsection{$d=2$}
\begin{figure}[t]
  \begin{center}
  \includegraphics*[width=7cm]{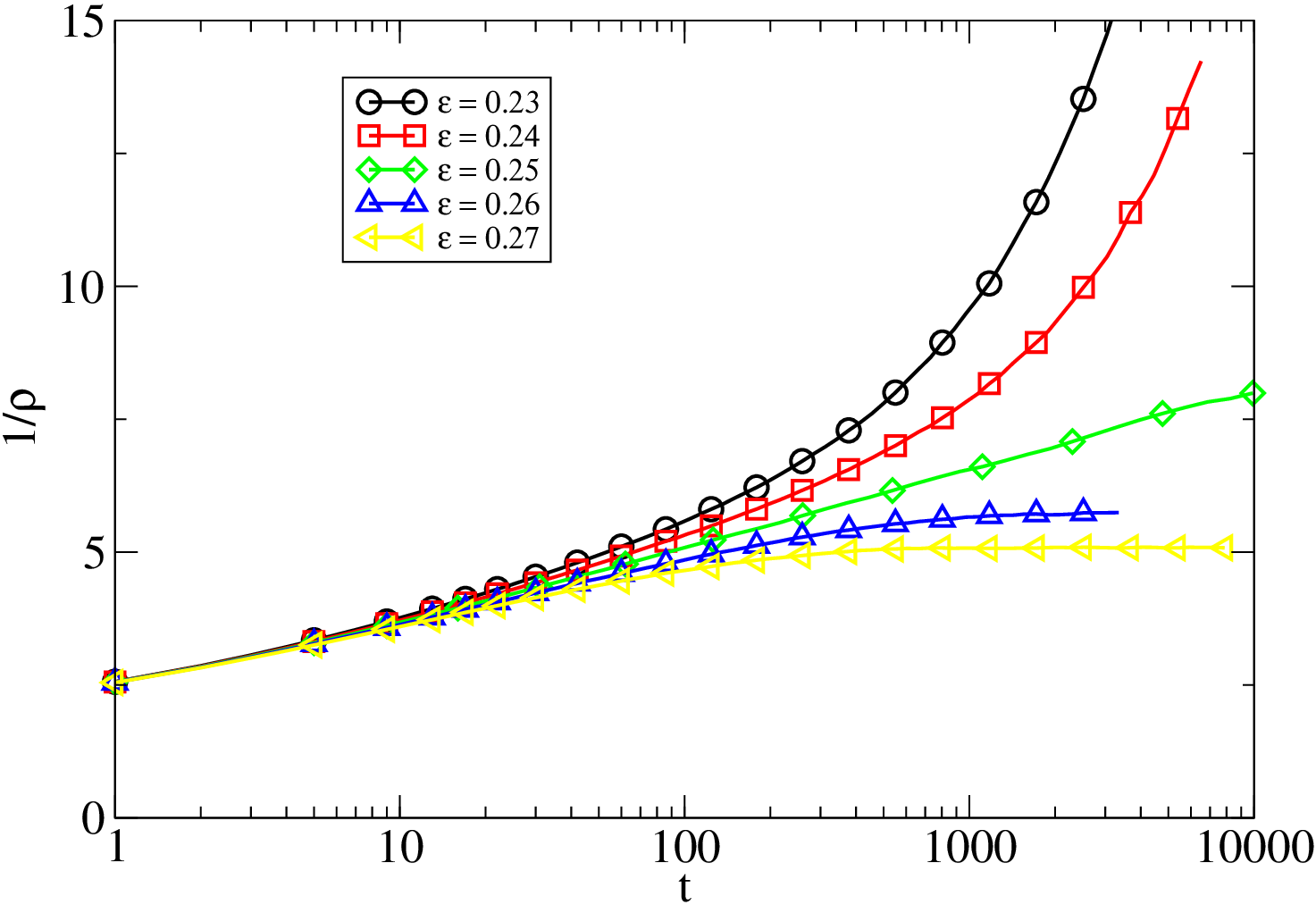}
\end{center}
\caption{(Color online)
Plot of the inverse of the density of active links $\rho(t)$ for different
values of $\ep$. Parameters: $q=4$, $\ep=1/4$,
$L=5000$.}
  \label{rhod=2}
\end{figure}

As in $d=1$, here $q$ is kept arbitrary.  We have performed numerical
simulations of the ordering dynamics of the $q$-voter model on a
square lattice of size $L \times L$ with $L=5000$, for several values
of $q>1$. In all cases, we find a transition separating a paramagnetic
phase for large $\ep$ from a ferromagnetic one at low $\ep$. In order
to investigate the nature of the transition we concentrate on the case
$q=4$, that in a fully connected graph yields non voter behavior.  In
Fig.~\ref{rhod=2}, we plot the temporal behavior of the inverse of
$\rho$, the fraction of active links in the system.
\begin{figure}[b]
  \begin{center}
    \includegraphics*[width=7cm]{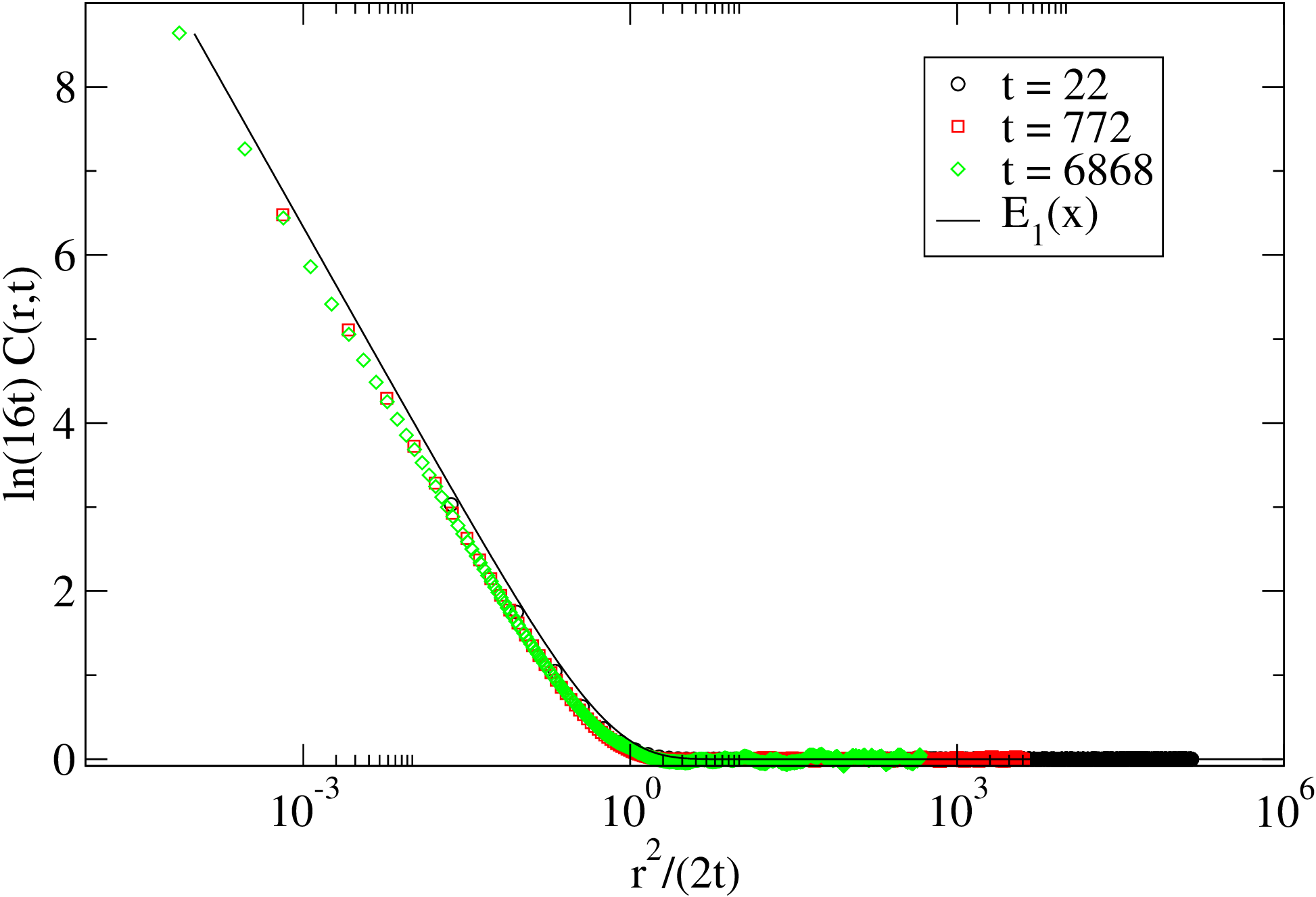}
  \end{center}
  \caption{(Color online) Collapse plot of the correlation function
    $C(r,t)$ for different times, according to Eq.~(\ref{scaling2d}).
    Parameters: $q=4$, $\ep=1/4$, $L=5000$.}
  \label{Scaling_C_2d}
\end{figure}
At the critical point $\ep=\ep_1=1/4$, $1/\rho$ grows logarithmically,
as expected for the voter universality class \cite{Frachebourg96}.
Analog results are found for other values of $q$
(data not shown).  Additional evidence proving that for $\ep=1/4$ the
$q$-voter model behaves exactly as the usual voter model is provided
by measuring the correlation function $C(r,t)$. From the exact
solution of the voter model in $d=2$~\cite{dornicthese} it turns out that
two different length scales are present in the system, leading to the
nonstandard scaling form 
\begin{equation}
  C(r,t) = \frac{1}{\ln(16t)}\tilde {f}[r^2/(2t)],
  \label{scaling2d}  
\end{equation}
where the $\tilde{f}(x) = E_1(x)$ is the exponential integral
function \cite{abramovitz}.  Figure~\ref{Scaling_C_2d} demonstrates
that Eq.~(\ref{scaling2d}) is nicely obeyed by numerical simulations
of the $q$-voter at the transition point for $q=4$.  This evidence,
further confirmed by the analysis of the exit probability,
showing a linear
behavior, leads to strong numerical confirmation that the scenario
predicted by Al Hammal \textit{et al.}~\cite{AlHammal05} is correct in
$d=2$: In finite dimensions, fluctuations renormalize the deterministic
potential so that the transition is in the GV class for
any value of $q$.
\begin{figure}[t]
  \begin{center}
    \includegraphics*[width=7cm]{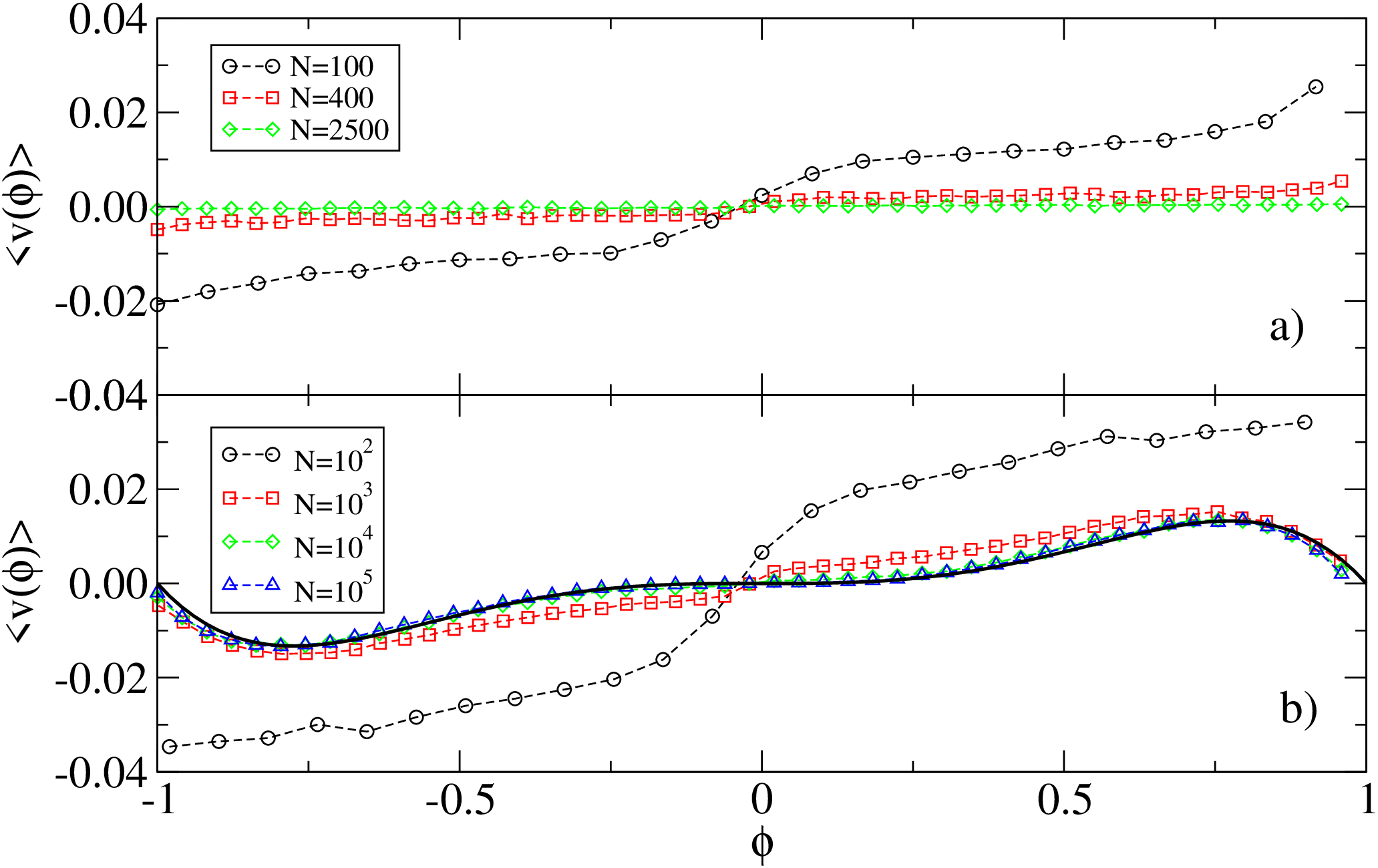}
  \end{center}
  \caption{(Color online) Numerical estimate of the average drift in
    the $q$-voter model, as a function of the magnetization.
    (a) Square lattice. (b) Fully connected network. In this last
    plot, the full line represents the expected theoretical value
    $v(\phi) \sim (1-\phi^2)\phi^3$.}
  \label{Drift_2d}
\end{figure}
This renormalization effect can be directly observed by measuring
numerically the drift term. From the Fokker-Planck equation, we can
obtain the equation for the time evolution of the
magnetization, $\av{\phi}$, namely \cite{Gardinerbook}
\begin{equation}
  \frac{d \av{\phi}}{d t} = \av{v(\phi)}.
\end{equation}

Therefore, a numerical evaluation of $d \av{\phi} / d t$ yields an
estimate of $\av{v(\phi)}$. In Fig.~\ref{Drift_2d} we plot the average
drift as a function of the magnetization for the $q$-voter
model in $d=2$ and for the mean-field fully connected case.  In the
latter, $\av{v(\phi)}$ shows a functional dependence compatible with
the theoretical expectation $v(\phi) \sim (1-\phi^2)\phi^3$ for
sufficiently large network size $N$.  In the $d=2$ case, on the other
hand, fluctuations are able to quickly cancel the drift term, inducing
thus an effective voter behavior in the limit of large $N$: $b<0$
renormalizes on large scales to $b=0$, as predicted in
\cite{AlHammal05}.

\section{Conclusions and discussion}
\label{conclusions}

We have introduced a non-linear variant of the voter model in which
the opinion of $q$ neighbors (with possible repetition) is taken into
account for a voter to change its own opinion. In particular, if all
his $q$ neighbors share the same state, an individual conforms to this
state; otherwise, if the $q$ neighbors do not agree, he flips with a
probability $\ep$.  Note that the model includes a noise effect
controlled by $\ep$; still voters are not allowed to break the
absorbing state condition and a consensus state remains indefinitely
so.  While the original definition of the model is meaningful only for
integer values of $q$, analytical generalization to arbitrary values
of $q$, with $q \in [0,\infty]$ is possible. In particular, after
taking the continuous limit for the transition rates, $q$ becomes a,
not necessarily integer, parameter.

We have studied the model analytically by applying a mean field analysis
based on the backward Fokker-Plank formalism and the Langevin approach
developed in Refs.~\cite{AlHammal05,Vazquez08}.  These two approaches
permit us to uncover the rich and variate phenomenology of the
$q$-voter model:
\begin{itemize}
\item For $ q=1$ the model reduces to the standard voter model with
  exit probability proportional to $x$ and average consensus time,
  starting from $x=1/2$, growing linear with system size.
\item For $ q=2$, the model coincides with the voter one if $\ep=1/2$.
  Instead, the system remains disordered for $\ep >1/2$, or it orders
  exponentially fast for $\ep <1/2$.  Therefore the $2$-voter model
  exhibits a ``generalized voter transition''.  Note that our results
  also clarify the behavior of the Sznajd model.  For $q=2$ and
  $\ep=0$ the $q$-voter model practically coincides with Sznajd model,
  at least in the formulation of Ref.~\cite{Slanina03}, according to
  which one has to select a pair of neighbors and, if they are in the
  same state, another neighbor of the pair is set in the same
  state. From this point of view, Sznajd model is just a ferromagnetic
  model in its ordered phase.
\item For $ q=3$ there is a voter like transition at $\ep=1/3$,
  separating two phases as those described for $q=2$ but, contrarily
  to the cases before, the flipping probabilities are non-linear: the
  $3$-voter model is an example belonging non-trivially to the
  generalized voter class.
\item For $2<q<3$ there is no voter transition. Instead the system
  experiences a sequence of two transitions. Starting with large
  values of $\ep$ and reducing it progressively, first the $Z_2$
  symmetry is broken (Ising transition) and afterward the system
  orders into an absorbing state at a directed-percolation like
  transition.
\item For $1<q<2$ and $q>3$ the mean-field approach predicts a
  non-voter transition, characterized by an exit probability with a
  Heaviside $\Theta$-function shape and a consensus time which
  increases with system size as $N^{1/2}$ rather than
  linearly. 
\end{itemize}

All these results have been verified in numerical simulations of the
model on a fully connected lattice.  On the other hand, in spatially
extended systems this last (third) scenario does not appear, as
predicted by Al Hammal \textit{ et al.}  \cite{AlHammal05}.  The
reason for this is, as we have numerically verified, that fluctuations
wash out the intermediate regime in which three stable states
exist. Indeed, as fluctuations can take the system from the origin to
any of the absorbing states but not the other way around, effectively,
on sufficiently large scales, the stable state at the origin plays no
role, and the system exhibits a single ordering transition in the
generalized-voter class.

Previous results were obtained under the rule that, among the $q$
neighbors involved in the dynamical step, each given neighbor may be
selected more than once.  If the possibility of repetition is
explictly forbidden, mean-field results clearly do not change, but in
finite dimensions variations are possible. We have performed numerical
simulations in $d=2$, showing that for $q<4$ no qualitative change
occurs: there is a GV transition between a disordered phase for large
$\ep$ and an ordered one for small $\ep$.  Things change only in the
special case $q=4$, where all neighbors of a site are considered. The
ordered phase disappears and for any value of $\ep$ a disordered state
is reached.

In summary, the $q$-voter model is a simple non-linear extension of
the voter model exhibiting a rich and interesting phenomenology and
illustrating how apparently innocuous changes in the microscopic
dynamics can lead to different types of collective phenomena, and in
particular to different paths to reach consensus.

\begin{acknowledgments}
  R. P.-S. acknowledges financial support from the Spanish MEC
  (FEDER), under project No. FIS2007-66485-C02-01, as well as
  additional support through ICREA Academia, funded by the Generalitat
  de Catalunya. M.A.M.  acknoledges support from the Junta de
  Andaluc\'{i}a project FQM--01505 and from Spanish MICINN--FEDER
  project FIS2009--08451. Useful discussions with H. Chat\'e and I.
  Dornic are kindly acknowledged.

\end{acknowledgments}

\end{document}